\documentstyle[12pt,epsf]{article}

\begin{document}

\title{Pair Creation and Evolution
      \\ of Black Holes in Inflation\thanks{based on a talk given by
        R.~Bousso at the Journ\'ees Relativistes 96 (Ascona,
        Switzerland, May 1996)}}
\author{{\sc Raphael Bousso}\thanks{\it R.Bousso@damtp.cam.ac.uk} \ 
  and {\sc Stephen W. Hawking}\thanks{\it S.W.Hawking@damtp.cam.ac.uk}
      \\[1 ex] {\it Department of Applied Mathematics and}
      \\ {\it Theoretical Physics}
      \\ {\it University of Cambridge}
      \\ {\it Silver Street, Cambridge CB3 9EW}
       }
\date{DAMTP/R-96/35}

\maketitle

\begin{abstract}

  We summarise recent work on the quantum production of black holes in
  the inflationary era. We describe, in simple terms, the Euclidean
  approach used, and the results obtained both for the pair creation
  rate and for the evolution of the black holes.

\end{abstract}

\pagebreak

\paragraph{Introduction} One usually thinks of black holes
forming through gravitational collapse, and so it seems that inflation
is not a good place to look for black holes, since matter is hurled
apart by the rapid cosmological expansion.  We will show, however,
that it is possible to get black holes in inflation through the
quantum process of pair creation~\cite{BouHaw95,BouHaw96}.  There are
two physical motivations that might lead us to expect this: First of
all, quantum fluctuations can be very large during inflation, which
leads to large density perturbations.  Secondly, in order to pair
create any objects, whether particles or black holes, one needs a
force to pull them apart.  Think of electron-positron pair creation:
unless there is a force pulling them apart, the virtual particles will
just fall back and annihilate.  But if they are in an external
electric field, the field pulls them apart and provides them with the
energy to become real particles.  Similarly, whenever one pair creates
black holes, one needs to do it on a background that will pull them
apart. This could be, for example, Melvin's magnetic universe, where
oppositely charged black holes are separated by the background
magnetic field, or a cosmic string, which can snap with black holes
sitting on the bare terminals, pulled apart by the string tension.
For the black holes we shall consider, the necessary force will be
provided by the rapid expansion of space during inflation. So this
expansion, which we naively thought would prevent black holes from
forming, actually enables pair creation.

\begin{figure}[htb]
\epsfxsize=\textwidth
\epsfbox{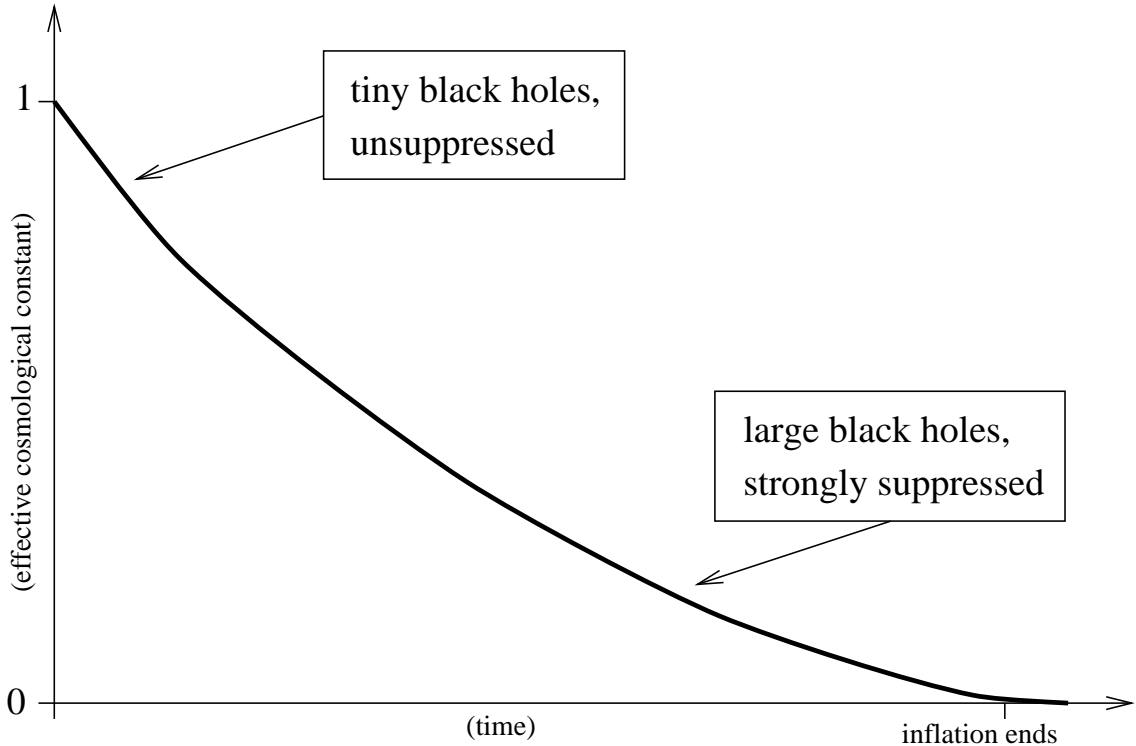}
\caption[Inflation]%
{\small\sl The classical evolution of the effective cosmological
  constant in a typical model of chaotic inflation. We have indicated
  qualitatively how the nucleation size and pair creation rate of
  black holes depend on the effective cosmological constant.}
\label{fig-inflation}
\end{figure}

\paragraph{Inflation} In quantum cosmology, one expects the
universe to begin in a phase called chaotic inflation. In this era the
evolution of the universe is dominated by the vacuum energy $V(\phi)$
of some inflaton field $\phi$. $V$ starts out at about the Planck
value, and then decreases slowly while the field rolls down to the
minimum of the potential. During this time the universe behaves like
de~Sitter space with an effective cosmological constant $\Lambda_{\rm
  eff} \approx V$ (see Fig.~\ref{fig-inflation}).
Like the scalar field, $\Lambda_{\rm eff}$
decreases only very slowly in time, and for the purposes of
calculating the pair creation rate, we can take $\Lambda$ to be
fixed~\cite{BouHaw95}.

\paragraph{Instanton method} An instanton is a Euclidean
solution of the Einstein equations, i.e., a solution with signature
(++++). Instantons can be used for the description of
non-per\-tur\-bat\-ive gravitational effects, such as spontaneous
black hole formation.  What follows is a kind of kitchen recipe for
this type of application.  We must consider two different spacetimes:
de~Sitter space without black holes (i.e., the inflationary
background), and de~Sitter space containing a pair of black holes.
For each of these two types of universes, we must find an instanton
which can be analytically continued to become this particular
Lorentzian universe.  The next step is to calculate the Euclidean
action $I$ of each instanton.  According to the Hartle-Hawking no
boundary proposal~\cite{HarHaw83}, the value of a wave function $\Psi$
is assigned to each universe. In the semi-classical approximation
$\Psi= e^{-I}$, neglecting a prefactor.  $P = |\Psi|^2 = e^{-2 I^{{\rm
      Re}}}$ is then interpreted as a probability measure for the
creation of each particular universe.  (Note that $P$ depends only on
the real part of the Euclidean action.)  The pair creation rate of
black holes on the background of de~Sitter space is finally obtained
by taking the ratio $\Gamma = P_{\rm BH}/P_{\rm no\, BH}$ of the two
probability measures.  One can also think of $\Gamma$ as the ratio of
the number of inflationary Hubble volumes containing black holes to
the number of empty Hubble volumes.

\begin{figure}[htb]
\epsfxsize=\textwidth
\epsfbox{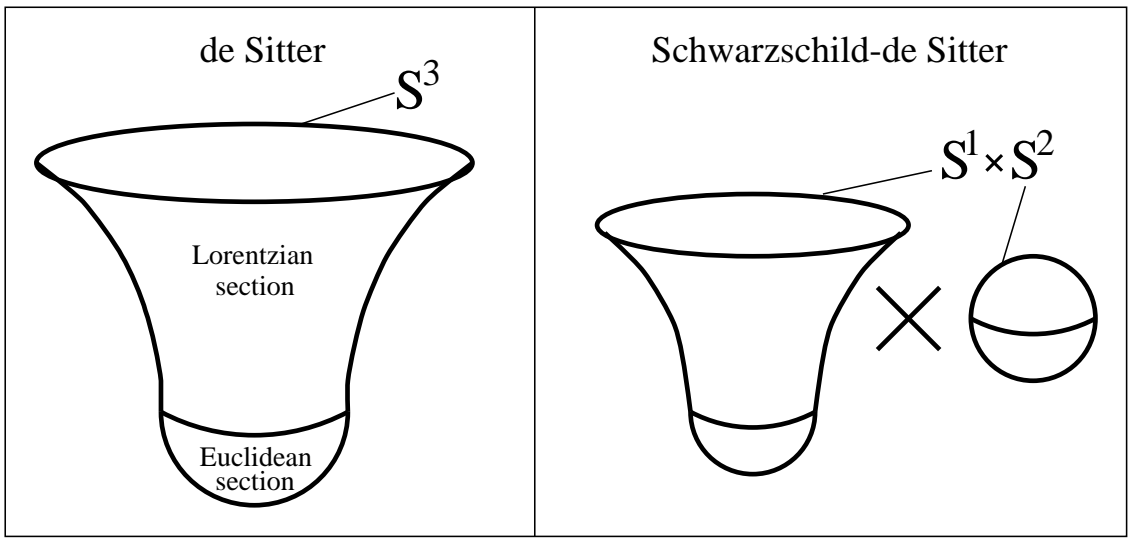}
\caption[Creation of de~Sitter and Schwarzschild-de~Sitter
universes]{\small\sl The creation of a de Sitter universe (left) can
  be visualised as half of a Euclidean four-sphere joined to a
  Lorentzian four-hyperboloid. The picture on the right shows the
  corresponding nucleation process for a de Sitter universe containing
  a pair of black holes. In this case the spacelike slices have
  non-trivial topology.}
\label{fig-tun}
\end{figure}

\paragraph{de~Sitter} We begin with the simpler of the two
spacetimes, an inflationary universe without black holes. In this case
the spacelike sections are round three-spheres. In the Euclidean de
Sitter solution, the three-spheres begin at zero radius, expand and
then contract in Euclidean time. Thus they form a four-sphere of
radius $\sqrt{3/\Lambda}$.  The analytic continuation can be
visualised (see Fig.~\ref{fig-tun})
as cutting the four-sphere in half, and then joining to it half the
Lorentzian de~Sitter hyperboloid, where the three-spheres expand
exponentially in Lorentzian time.  The real part of the Euclidean
action for this geometry comes from the Euclidean half-four-sphere
only: $I^{\rm Re}_{\rm no\, BH} = - 3\pi/2\Lambda$.  Correspondingly,
the probability measure for de~Sitter space is
\begin{equation}
P_{\rm no\, BH} = \exp \left( \frac{3\pi}{\Lambda} \right).
\end{equation}

\begin{figure}[htb]
\epsfxsize=\textwidth
\epsfbox{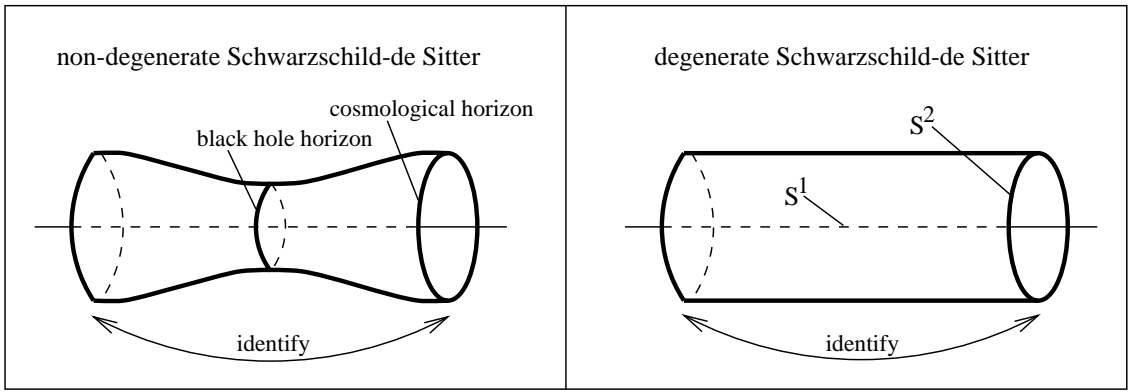}
\caption[Spacelike sections of Schwarzschild-de~Sitter space]%
{\small\sl The spacelike slices of Schwarzschild-de~Sitter space have
  the topology $S^1 \times S^2$. In general (left), the size of the
  two-sphere varies along the one-sphere. If the black hole mass is
  maximal, however, all the two-spheres have the same size (right).
  Only in this case is a smooth Euclidean solution admitted.}
\label{fig-space-secs}
\end{figure}

\paragraph{Schwarzschild-de Sitter} Now we need to go through
the same procedure with the Schwarzschild-de~Sitter solution, which
corresponds to a pair of black holes immersed in de~Sitter space.  The
spacelike sections in this case have the topology $S^1 \times S^2$.
This can be seen by the following analogy: Empty Minkowski space has
spacelike sections of topology ${\rm {\bf R}}^3$. Inserting a black
hole changes the topology to $S^2 \times {\rm {\bf R}}$. Similarly, if
we start with de~Sitter space (topology $S^3$), inserting a black hole
is like punching a hole through the three-sphere, thus changing the
topology to $S^1 \times S^2$.  In general, the radius of the $S^2$
varies along the $S^1$. The maximum two-sphere corresponds to the
cosmological horizon, the minimum to the black hole horizon. This is
shown in Fig.~\ref{fig-space-secs}.

What we need is a Euclidean solution that can be analytically
continued to contain this kind of spacelike slice.  It turns out that
such a smooth instanton does not exist in general for the Lorentzian
Schwarzschild-de~Sitter spacetimes. The only exception is the
degenerate case, where the black hole has the maximum possible size,
and the radius of the two-spheres is constant along the $S^1$ (see
Fig.~\ref{fig-space-secs}). The corresponding Euclidean solution is
just the topological product of two round two-spheres, both of radius
$1/\sqrt{\Lambda}$~\cite{GinPer83}.  It can be analytically continued
to the Lorentzian Schwarzschild-de~Sitter solution by cutting one of
the two-spheres in half, and joining to it the 2-dimensional
hyperboloid of $1+1$ dimensional Lorentzian de~Sitter space, as shown
in Fig.~\ref{fig-tun}.  In the Lorentzian regime the $S^1$ expands
exponentially, while the two-sphere just retains its constant radius.
Thus the Euclidean approach predicts the size with which the black
holes will be nucleated:
\begin{equation}
r_{\rm BH} = \sqrt{\frac{1}{\Lambda}}.
  \label{eq-bh-radius}
\end{equation}
The real part of the Euclidean action for this instanton is given by
$I^{\rm Re}_{\rm BH} = - \pi/\Lambda$, and the corresponding
probability measure is
\begin{equation}
P_{\rm BH} = \exp \left( \frac{2\pi}{\Lambda} \right).
\end{equation}

\paragraph{Pair creation rate}
Now we can take the ratio of the two probability measures, and obtain
the pair creation rate: 
\begin{equation}
\Gamma = \exp \left( -\frac{\pi}{\Lambda} \right).
\end{equation}
Let us interpret this result.  The cosmological constant is positive
and no larger than order unity in Planck units.  This means that black
hole pair creation is suppressed.  When $\Lambda \approx 1$ (early in
inflation), the suppression is week and one can get a large number of
black holes.  However, by Eq.~(\ref{eq-bh-radius}), they will be very
small (Planck size).  For smaller values of $\Lambda$ (which are
attained later in inflation), the black holes would be larger, but
their creation becomes exponentially suppressed (see
Fig.~\ref{fig-inflation}).  This result, which was obtained from the
no boundary proposal, is physically very sensible.

\paragraph{Tunnelling proposal} According to Vilenkin's
tunnelling proposal~\cite{Vil86}, the wave function is given by
$e^{+I}$, rather than $e^{-I}$. If we tried to apply this prescription
to our problem, the signs would get reversed in all the exponents, and
we would get the inverse result for $\Gamma$. Thus black hole creation
would be enhanced, rather than suppressed. Even worse, the bigger the
black holes were, the more likely they would be to nucleate. As a
consequence, de~Sitter space would be catastrophically unstable. This
prediction is obviously absurd.  Thus, the consideration of
cosmological black hole pair creation provides strong evidence in
favour of the no boundary proposal.

\paragraph{Classical evolution} What happens to black holes
that have been pair created during inflation?  In the above instanton
solution they would just retain their constant size $r_{\rm BH} =
1/\sqrt{\Lambda}$.  But in this case it is important to take into
account that during inflation, the effective cosmological constant
isn't fixed, but decreases slowly.  With this correction, the black
hole radius during inflation is given by $r_{\rm BH} =
1/\sqrt{\Lambda_{\rm eff}}$.  As the inflaton field rolls down,
$\Lambda_{\rm eff}$ decreases, and the black hole grows slowly,
becoming quite large by the end of inflation.  This growth can be
explained by the First Law of black hole mechanics, which states that
the increase in a black hole's horizon area, multiplied by its
temperature, is equal to four times the increase in its mass. The mass
increase comes from the flux of energy-momentum of the inflaton field
across the black hole horizon, as the field rolls down.

\paragraph{Quantum evolution} There are some quantum effects
on the evolution which we have not yet taken into account. It is well
known that both the black hole and the cosmological horizon emit
radiation. The temperature of each horizon is approximately
proportional to its inverse radius.  In the instanton solution the
radii of the two horizons will be equal, and, therefore, also their
radiation rates.  The black hole loses as much mass due to Hawking
radiation as it gains from the incoming cosmological radiation, and it
would seem to be stable.  Because of quantum fluctuations, however,
the radius of the two-spheres will vary slightly along the one-sphere.
Then the black hole will be smaller and hotter than the cosmological
horizon.  It starts to lose mass and evaporates.  Only if it was
created very late in inflation would it be massive and cold enough to
grow classically and survive into the radiation era.  But such black
holes are highly suppressed.  The tiny, hot black holes created early
in inflation will all evaporate immediately.  Therefore there will be
no significant number of neutral black holes after inflation ends.

\paragraph{Magnetically charged black holes} There also are
instantons that correspond to the creation of magnetically charged
black holes.  Such black holes cannot evaporate altogether, because
there are no magnetically charged particles they could radiate.
Therefore they are still around today.  A detailed calculation shows,
however, that they are so suppressed, and so strongly diluted by the
inflationary expansion, that there won't even be a single charged
primordial black hole in the observable universe.  (This is a sensible
prediction, since we don't observe any.)  In dilatonic theories of
inflation, however, their number could be significantly larger; this
is currently being investigated.

\paragraph{Summary} Semi-classical calculations indicate that
tiny black holes are plentifully produced at the Planck era. The
creation of larger black holes is exponentially suppressed. During
inflation, the black holes can grow classically, but will mostly
evaporate due to quantum effects. Magnetically charged black holes
cannot evaporate, but their number today is exponentially small.
Generally, in the context of cosmological pair creation of black
holes, the no boundary proposal gives physically sensible results,
while the tunnelling proposal does not seem to be applicable.

\end{document}